\documentclass[version=preprint]{iacrcc}
\usepackage{headNGCiC}
\usepackage{import}

\usepackage[normalem]{ulem}
\usepackage{multirow}
\license{CC-by}
\title[running  = {Why cut-and-choose quantum state verification cannot be both efficient and secure},
      ]{Why cut-and-choose quantum state verification cannot be both efficient and secure}
\addauthor[orcid    = {0000-0003-1309-9786},
           inst     = {1},
           email    = {f.wiesner@tu-berlin.de},
           surname  = {Wiesner}
          ]{Fabian Wiesner}

\addauthor[orcid   = {0009-0009-8270-627X},
           inst    = {1},
           email   = {ziad.chaoui@tu-berlin.de},
           surname = {Chaoui},
          ]{Ziad Chaoui}
\addauthor[orcid   = {0009-0006-6355-1581},
           inst    = {2},
           email   = {diana-maria.kessler@taltech.ee},
           surname = {Kessler},
          ]{Diana Kessler}
\addauthor[orcid   = {0000-0002-4662-149X},
           inst    = {1},
           email   = {anna.pappa@tu-berlin.de},
           surname = {Pappa},
          ]{Anna Pappa}
\addauthor[orcid   = {0000-0002-8919-343X},
           inst    = {3},
           email   = {martti.karvonen@ucl.ac.uk},
           surname = {Karvonen},
          ]{Martti Karvonen}
\authorrunning{F.Wiesner et al.}

\addaffiliation[country={Germany},ror = {03v4gjf40}]{Technische Universität Berlin}
\addaffiliation[country={Estonia}, ror = {0443cwa12}]{Tallinn University of Technology}
\addaffiliation[country={UK},ror = {02jx3x895}]{University College London}

\begin{document}

\maketitle

\begin{abstract}
\noindent 
Quantum state verification plays a vital role in many quantum cryptographic protocols, as it allows the use of quantum states from untrusted sources. While some progress has been made in this direction, the question of whether the most prevalent type of quantum state verification, namely cut-and-choose verification, can be efficient and secure, is still not answered in full generality. 
In this work, we show a fundamental limit for quantum state verification for all cut-and-choose approaches used to verify arbitrary quantum states. We provide a no-go result showing that the cut-and-choose techniques cannot lead to quantum state verification protocols that are both efficient in the number of rounds and secure. 
We show this trade-off for stand-alone and composable security, where the scaling of the lower bound for the security parameters renders cut-and-choose quantum state verification effectively unusable.
\end{abstract}
\begin{textabstract}
Quantum state verification plays a vital role in many quantum cryptographic protocols, as it allows using quantum states from an untrusted source. While some progress has been made in this direction, the question of whether the most prevalent type of quantum state verification, namely cut-and-choose verification, can be efficient and secure, is still not answered in full generality. 
In this work, we show a fundamental limit for quantum state verification for all cut-and-choose approaches used to verify arbitrary quantum states. We provide a no-go result showing that the cut-and-choose techniques cannot lead to quantum state verification protocols that are both efficient and secure. 
We show this trade-off for stand-alone and composable security, where the scaling of the lower bound for the security parameters renders cut-and-choose quantum state verification effectively unusable.
\end{textabstract}

\section{Introduction}
For much of cryptography's history, security has been assumed but not proven. Even today, we rely on protocols whose security proofs are based on conjectured hardness assumptions \cite{Katz_Lindell}. In the comparably young field of quantum cryptography, many protocols claim provable security under the assumption that the devices used in these protocols are trustworthy. Although they offer a real advantage in tackling modern cryptographic challenges \cite{Quantum_crypto_rev,Pirandola:20}, they often come with two caveats:
    \begin{enumerate}
        \item Quantum hardware is expensive and difficult to operate and maintain. This is particularly true for quantum computers and their main building blocks, such as implementations of entangling gates \cite{Preskill_2018}.
        \item The devices might not be trustworthy. To assume otherwise might in fact be a very strong assumption; someone untrusted could be operating the device, or there could be a hardware-based attack that leaks important information, as was done in the past for quantum key distribution systems \cite{lydersen:hacking}.
    \end{enumerate}

   Interestingly, these two issues are connected. Indeed, one way to address the first issue is to delegate some complex tasks to other parties while ensuring they execute them as required. Quantum correlations provide a way to check that the operations and tasks at hand are executed correctly. In the most general case, this is done through a framework called ``Device-Independence'' \cite{Ac_n_2007}, where the parties involved in the protocol can verify that the operations performed are correct, without putting any trust in the hardware. 
    \\
    In this paper, we focus on one specific task: quantum state verification. Quantum state verification is crucial to a variety of applications, such as network parameter estimation \cite{shettell_PrivateNetworkParameter_2022}, in which certain parties (the clients) cannot prepare a required quantum state themselves\footnote{Note that if the clients could prepare the required quantum state, they would not query the untrusted source and would not use quantum state verification} and therefore need to query a potentially dishonest source.  
    Usually, in quantum state verification protocols, the untrusted source prepares quantum states and distributes them among the clients who are sometimes considered honest. If the source is honest, it always prepares the target state, i.e. the state the clients desire to hold, and the clients accept the result. However, if the source is dishonest, it might not always send the target state, and the clients should ideally reject it. Due to the no-cloning theorem, clients cannot simply measure and use the quantum states sent by the source. Hence, some other form of verification is necessary. A typical way to verify quantum states is for the source to send several copies of the state, some of which are then measured by the clients. If sufficiently many measurements match the expected state, the clients are convinced the source is honest.
    This \textit{cut-and-choose} type of verification is used in many applications, such as anonymous conference key agreement \cite{dejong:anonymousCKA}, network parameter estimation \cite{shettell_PrivateNetworkParameter_2022}, and blind quantum computing \cite{hayashi_blind_2015}. With the emergence of quantum technologies and the effort to build a secure quantum internet \cite{wehner_internet_2018}, these applications are of increasing importance. For cut-and-choose quantum state verification to be a viable subroutine to all these protocols, we expect it to be composably secure. While composable security is usually defined with respect to a certain framework (e.g. abstract cryptography), one can prove negative results in a more general way. Indeed, we present a no-go result for such verification techniques, which is valid for composable security (independent of the framework) as well as for stand-alone security.

\subsection{Our contribution and related work}
Many protocols implement quantum state verification for different types of states, e.g. \cite{hypergraph,many_qubit,Pappa_2012,Unnikrishnan_2022}. However, all protocols we are aware of solely rely on cut-and-choose and suffer from the same efficiency vs. security trade-off: a quantum state verification protocol cannot be both secure and efficient. We investigate this trade-off in a more general setting and prove the following no-go result for quantum state verification.
\begin{theorem}[Main result (informal)]\label{thm:informal}
    Let $\pi$ be a cut-and-choose protocol for quantum state verification in which the clients output the state without performing any map on it, if they evaluate the source's behavior as honest.
    At least one of the following statements about $\pi$ with security parameter $\lambda$ is false:
    \begin{enumerate}
        \item $\pi$ rejects the target state with a probability negligible in $\lambda$.
        \item If the source is dishonest, either the probability to accept or the distance to the target state is negligible in $\lambda$.
        \item The number of rounds $N$ is polynomial in $\lambda$.
    \end{enumerate}
    Regarding composable security, we find with $\varepsilon_H$ being the distinguishability to the idealized process if the source is honest and $\varepsilon_D$ if it is not
    \begin{align*}
        \varepsilon_H+\varepsilon_D\geq \frac{\sqrt{\eta_1}}{4\sqrt{N}},
    \end{align*}
    where $\eta_1$ is the highest eigenvalue of the target state (i.e. $1$ if the target state is pure).\\
Regarding stand-alone security, we find with $\kappa_H$ being the fidelity between the honest output and $\phi$ and  $\kappa_D$ being the fidelity of the dishonest output and $p\phi \oplus (1-p)$\footnote{Note, that we consider the abort probability $1-p$ in a one-dimensional space using the direct sum, see the Section \ref{sec:QITPrelim} for more details.} maximized over $p$
 \begin{align*}
        \kappa_H+\kappa_D\geq \frac{1}{7N} .
    \end{align*}
\end{theorem}

We state and prove these results more formally in Section \ref{sec:ng-iid}. We show that for a generic cut-and-choose protocol described in Algorithm \ref{proto:simple} 
the inequalities of Theorem \ref{thm:informal} hold. These inequalities imply a trade-off between correctness, security, and efficiency: A protocol with high correctness, that is with high $\varepsilon_H$ or $\kappa_H$, has low security, i.e. low $\varepsilon_D$ or $\kappa_D$, and vice versa. The efficiency of the protocol is given by the number of rounds $N$, and we can see that the lower bounds for the sums of correctness and security scale inversely with the round numbers. 
Similar trade-offs have been proven in other works before in the hypothesis testing framework\footnote{See \cite{QSV-review} for a review on quantum state verification focused on the hypothesis testing approach.}. 
More specifically, the previous works showing similar trade-offs \cite{Bound1,Bound2} consider a protocol to be $(\varepsilon,\delta)$-secure, if the probability that the client accepts the behavior of the source is upper bounded by $\delta$ when the deviation from the honest behavior is intolerable.
This deviation is defined to be intolerable if the fidelity between the average output state of the protocol $\rho$ and the target state $\phi$ is upper bounded by $1-\varepsilon$, where the average output state is obtained by the composition of any attack of the source and the protocol of the client, i.e.
\begin{align*}
    Pr[{\rm accept}|F(\rho,\phi)\leq 1-\varepsilon]\leq \delta.
\end{align*}
Intuitively, this means that the probability that the client outputs a state that is too different from the target state should be low for all attacks the source might use. Using this definition and the assumption that the acceptance probability if the source is honest is $1$, the authors of \cite{Bound1} and \cite{Bound2} showed that for a quantum verification protocol for any pure target state $\phi$ with a fixed number of $N+1$ rounds, it holds that\footnote{Note that dividing by $\ln(1-c\varepsilon)$ changes the direction of the inequality since $\ln(1-c\varepsilon)\leq 0$.}
\begin{align*}
    Pr[{\rm accept}|F(\rho,\phi)\leq 1-\varepsilon] &= \left(1-c\varepsilon\right)^N \leq \delta\\
    \Rightarrow N \geq \frac{\ln(\delta)}{\ln(1-c\varepsilon)}
\end{align*}
where $c = 1$ in \cite{Bound1} and $c\leq 1$ is a constant in \cite{Bound2} which depends on the verification strategy.\\ 
Both, \cite{Bound1} and \cite{Bound2}, assume that the clients perform single-round tests, i.e., do not use coherent measurements, although in \cite{Bound1} the authors argue that this is not a restriction. 
However, our work differs from \cite{Bound1, Bound2} in many aspects. First, the assumptions differ: we allow for \emph{coherent measurements}, we derive a bound for \emph{mixed target states} as well, do not require \emph{perfect correctness}, and, very importantly, we do not consider a \emph{fixed number of rounds}. While we note that single-state measurements suffice to optimally distinguish pure states and that one could use Uhlmann's theorem to derive a bound for mixed states as well, a fixed round number is a strong assumption for the protocol. Especially if the client does not use coherent but uncorrelated measurements, one can randomize the number of rounds, e.g., as in \cite{Pappa_2012}, to obtain protocols outside of the scope of previous results. Indeed, there is an intuitive attack against protocols with a fixed round number: An attacker always sends the target state except for one round, for which it sends an orthogonal state, guessing that this is the output round. Such an intuitive attack is generally not available if the round number is randomized or the attacker is i.i.d.-restricted. 

However, our perspective on the topic also differs from that of previous works. In contrast to \cite{Bound1,Bound2}, we do not use the hypothesis testing framework, which is not common in many areas of quantum cryptography, despite being useful for quantum state verification.
We argue, in line with \cite{Yehia_2021,colisson2024graph}, that quantum state verification should be viewed as a building block of larger protocols and hence investigate composability as well. 
Because of this difference, we developed a framework-agnostic proof technique by refuting an implication of security that all composable security frameworks share, i.e. we find a lower bound on the trace distance between ideal and real output states of a general quantum state verification protocol, which implies a no-go result in the different frameworks for composable cryptography. For both types, composable and stand-alone security, we present i.i.d. attacks that break cut-and-choose quantum state verification. While we find, for a fixed round number, that the intuitive attack is optimal for stand-alone security, our presented i.i.d. attack achieves a higher violation of composable security. We expect that our proofs can be adapted for other functionalities for which hypothesis testing is less common.\\
So while a direct comparison of the different bounds is not reasonable due to the differences in the security definitions and assumptions, a summary and comparison of the assumptions with the previous results is presented in Table \ref{table:comparison}.
\begin{table}[h]
    \centering
    \resizebox{\textwidth}{!}{
    \begin{tabular}{l|l|l|l}
                   & Security type          & Bound & Assumptions\\\hline
        \cite{Bound1}   & Hypothesis testing    & $N \geq \frac{\ln(\delta)}{\ln(1-\varepsilon)}$ & 1,2,3,4\\\hline
        \cite{Bound2}   & Hypothesis testing    & $N \geq \frac{\ln(\delta)}{\ln(1-c\varepsilon)}$ & 1,2,3,4\\\hline
        This Work, Lem. \ref{lemma:equal}, \ref{lemma:genlowbound} and Thm. \ref{theorem:optimalFidAttack}   & Fidelity-based    & $\varepsilon_H+\varepsilon_D \geq \nicefrac{1}{N+1}$ & 1,2,4 \\\hline
        This Work, Thm. \ref{theorem:FidGen} & Fidelity-based    & $\varepsilon_H+\varepsilon_D \geq \nicefrac{1}{7N}$ & - \\\hline
        This Work, Thm. \ref{theorem:composGen}& Composable    & $\varepsilon_H+\varepsilon_D \geq \nicefrac{1}{4\sqrt{N}}$ & -
    \end{tabular}}
    \caption{Summary of results regarding the trade-off. Note that while \cite{Bound2} considers arbitrary attacks, in \cite{Bound1} only i.i.d. attacks are possible.\\
    Assumptions about the protocols as discussed above: 1) Fixed round number, 2) pure target state, 3) perfect correctness, 4) No coherent measurements for verification.}\label{table:comparison}
\end{table}

Finally, our results provide bounds for self-testing as well \cite{Self-testing-review}. Self-testing is slightly different from quantum state verification, since there is a single client that does not trust any of their devices, including the preparation and measurement apparatus. Self-testing can therefore be seen as a stricter case of quantum state verification. Hence, any attack on quantum state verification implies an attack on self-testing.

\subsection{Structure}
Our work is structured as follows: In Section \ref{sec:QITPrelim}, we first present preliminaries that we need for our security proofs. In Section \ref{sec:ng-iid}, we provide first the no-go result for a fidelity-based security definition and then for generic composable security with an i.i.d. restriction for the attacker. In Section \ref{sec:ng-non-iid}, we investigate optimal attacks outside of the i.i.d. setting. Finally, we discuss open questions and possible implications of our work in Section \ref{sec:disc}. In the appendix, we present a generalization of our no-go result for protocols with a probabilistic round number, and the security proof for a specific protocol, which provides guidance regarding the tightness of the bounds we prove and the advantage of more advanced attacks.

\section{Preliminaries}
    \label{sec:QITPrelim}
    In the following we present some preliminaries -- mainly on quantum information theory -- that we need for our security analyses in the subsequent sections.  

    We denote by $D(\X)$ the space of density operators on the Hilbert space $\X$. For a density operator $\rho\in D(\X)$, we define the \emph{trace norm} to be
     $\|\rho\|_1\coloneqq\Tr(\sqrt{\rho\rho\D})$. $P(\X)$ is the space of all positive semidefinite operators, and we define a (binary) measurement to be a function of the form $\mu:\{0,1\}\to P(\X)$, satisfying $\mu(0)+\mu(1)=\mathbbm{1}_\X$. For a density operator $\rho\in D(\X)$, $\braket{\mu(b)}{\rho}\coloneq \Tr(\mu(b)\D \rho)$ is then the probability to obtain measurement outcome $b\in \{0,1\}$ when measuring $\rho$ with $\mu$.
     The trace distance yields a bound on the achievable distinguishing advantage of a measurement between two density operators given by the Holevo-Helstrom Theorem.
    
    \begin{theorem}[Holevo-Helstrom Theorem]\label{thm:holevo_helstrom}
    Let $\rho_0, \rho_1\in D(\X)$ be density operators, and let $\lambda\in [0, 1]$. For any measurement $\mu:\{0, 1\}\rightarrow P(\X)$  it then holds
    \begin{equation}\label{eq:holevo_helstrom}
        \lambda \braket{\mu(0)}{\rho_0}+ (1-\lambda) \braket{\mu(1)}{\rho_1}\leq \frac{1}{2}+\frac{1}{2}\|\lambda\rho_0-(1-\lambda)\rho_1\|_1.
    \end{equation}
    Moreover there exists a projective measurement $\mu:\{0, 1\}\rightarrow P(\X)$ for which equality is achieved in (\ref{eq:holevo_helstrom}).
    \end{theorem}
    
    To see that this actually gives a bound on the distinguishing advantage, we set $\lambda=\frac{1}{2}$ in (\ref{eq:holevo_helstrom}) and we obtain
    \begin{align}
            &\frac{1}{2} \braket{\mu(0)}{\rho_0}+  \frac{1}{2} \braket{\mu(1)}{\rho_1}\leq \frac{1}{2}+ \frac{1}{4}\|\rho_0-\rho_1\|_1\nnn
            \Leftrightarrow& (\braket{\mu(1)}{\rho_1}-1) + \braket{\mu(0)}{\rho_0} = \braket{\mu(0)}{\rho_0} -\braket{\mu(0)}{\rho_1}\leq \frac{1}{2}\|\rho_0-\rho_1\|_1.\label{eq:dist_adv_pre}
    \end{align}
    If $\braket{\mu(0)}{\rho_0}\geq\braket{\mu(0)}{\rho_1}$ holds, then \eqref{eq:dist_adv_pre} gives a bound on the distinguishing advantage $|\braket{\mu(0)}{\rho_0} -\braket{\mu(0)}{\rho_1}|$ which is the absolute value of the difference of the probabilies of the oucome $0$. If, however, $\braket{\mu(0)}{\rho_0}<\braket{\mu(0)}{\rho_1}$, we define the measurement operator $\gamma(0)\coloneqq \mathbbm{1}_{\X}-\mu(0)$ and find $\braket{\gamma(0)}{\rho_0}\geq\braket{\gamma(0)}{\rho_1}$. Since \eqref{eq:holevo_helstrom} holds for every measurement we then find
    \begin{align*}
        \braket{\gamma(0)}{\rho_0} -\braket{\gamma(0)}{\rho_1} =  (1-\braket{\mu(0)}{\rho_0}) -(1-\braket{\mu(0)}{\rho_1}) = \braket{\mu(0)}{\rho_1}-\braket{\mu(0)}{\rho_0}\leq \frac{1}{2}\|\rho_0-\rho_1\|_1
    \end{align*}
    Hence, either way it holds 
    \begin{align}\label{eq:dist_adv}
        \left|\braket{\mu(0)}{\rho_0} -\braket{\mu(0)}{\rho_1}\right|\leq \frac{1}{2}\|\rho_0-\rho_1\|_1.
    \end{align}
    
    Another important quantity is the fidelity. The fidelity between two density operators $\rho_0$, $\rho_1$ is given by\footnote{Note that other authors define the square root of this expression as the fidelity.}
    \begin{equation*}
    F(\rho_0, \rho_1)\coloneqq\Tr\left(\sqrt{\sqrt{\rho_0}\rho_1\sqrt{\rho_0}}\right)^2.
    \end{equation*}
    Although the fidelity is not a metric, it allows to quantify how close or similar two density operators are: the higher the fidelity, the closer the states.  We also define the infidelity to be $1-F(\rho_0,\rho_1)$. For all pure states $\rho_0=\ketbra{\psi_0}$ and $\rho_1=\ketbra{\psi_1}$ it holds that
    \begin{equation}\label{eq:pure_state_fid}
        F(\rho_0,\rho_1) = \bra{\psi_0}\rho_1\ket{\psi_0} = |\braket{\psi_0}{\psi_1}|^2,
    \end{equation}
    where the first equality holds even if $\rho_1$ is not pure.
    Further, we have the following properties for all density operators $\rho_0, \sigma_0\in D(\X)$, $\rho_1,\sigma_1\in D(\Y)$, and $\lambda\geq 0$
    \begin{align}\label{eq:fidOtimes}
     F(\rho_0\otimes \rho_1, \sigma_0\otimes\sigma_1)&=F(\rho_0,\sigma_0)F(\rho_1, \sigma_1).\\\label{eq:fidOplus}
        \sqrt{F(\rho_0\oplus \rho_1,\sigma_0\oplus \sigma_1)} &= \sqrt{F(\rho_0,\sigma_0)}+\sqrt{F(\rho_1,\sigma_1)}\\\label{eq:fidLin}
        F(\lambda \rho_0,\sigma_0) &= \lambda F(\rho_0,\sigma_0),
    \end{align}
    where $R\oplus Q$ denotes the direct sum for linear operators, i.e.
    \begin{align*}
        R \oplus Q = \left(\begin{matrix}
            R_{1,1} & ... & R_{1,b}\\
            \vdots & \vdots & \vdots\\
            R_{a,1} & ... & R_{a,b}\\
        \end{matrix}\right)\oplus\left(\begin{matrix}
            Q_{1,1} & ... & Q_{1,d}\\
            \vdots & \vdots & \vdots\\
            Q_{c,1} & ... & Q_{c,d}\\
        \end{matrix}\right) \coloneqq 
        \left(\begin{matrix}
            R_{1,1} & ... & R_{1,b} & 0 & ... & 0\\
            \vdots & \vdots & \vdots & \vdots & \vdots & \vdots \\
            R_{a,1} & ... & R_{a,b}& 0 & ... & 0\\
            0 & ... & 0 & Q_{1,1} & ... & Q_{1,d}\\
            \vdots & \vdots & \vdots & \vdots & \vdots & \vdots \\
            0 & ... & 0 & Q_{c,1} & ... & Q_{c,d}\\
        \end{matrix}\right).
    \end{align*}
    The Fuchs-van de Graaf inequalities link the trace distance to the fidelity~\cite{watrous_2018}. 
    \begin{theorem}[Fuchs-van de Graaf Inequalities] \label{thm:FuchsVDG}Let $\rho_0, \rho_1\in D(\X)$ be density operators, it  holds that
    \begin{align}
       &1-\sqrt{F(\rho_0, \rho_1)}\leq \frac{1}{2}\|\rho_0-\rho_1\|_1\leq
    \sqrt{1-F(\rho_0, \rho_1)}\label{eq:FvdG_1},\\
    &\left(1-\frac{1}{2}\|\rho_0-\rho_1\|_1\right)^2\leq F(\rho_0, \rho_1)\leq 1-\frac{1}{4}\|\rho_0-\rho_1\|_1^2.\label{eq:FvdG_2}
    \end{align}
    \end{theorem}
    Using these inequalities and the properties of fidelity, we can easily derive a simple bound on $k$-tuples of density operators $\{(\rho_i,\sigma_i)\}_{i=1}^k$:
   \begin{equation}\label{eq:Multi_copy_dist}
            \frac{1}{2}\left\|\bigotimes_{i=1}^k\rho_i-\bigotimes_{i=1}^k\sigma_i\right\|_1\stackrel{(\ref{eq:FvdG_1})}{\leq}\sqrt{1-F\left(\bigotimes_{i=1}^k\rho_i, \bigotimes_{i=1}^k \sigma_i\right)}
            \stackrel{(\ref{eq:fidOtimes})}{=}\sqrt{1-\prod_{i=1}^k F\left(\rho_i, \sigma_i\right)}.
    \end{equation}
    For pure states $\rho=\ketbra{\psi}$, $\sigma=\ketbra{\phi}$, it holds that \begin{equation}\label{eq:trace_pure}
         \|\ketbra{\psi}-\ketbra{\phi}\|_1 = 2\sqrt{1-|\braket{\psi}{\phi}|^2},
     \end{equation}which implies for $\rho_i=\ketbra{\psi_i}$, $\sigma_i=\ketbra{\phi_i}$
        \begin{align}\label{eq:pureState}
            \frac{1}{2}\left\|\bigotimes_{i=1}^k\rho-\bigotimes_{i=1}^k\sigma_i\right\|_1 = \sqrt{1-\prod_{i=1}^k\left|\braket{\psi_i}{\phi_i}\right|^2}.
        \end{align}
%\vspace{-2em}\sectionbreak[* * *]\vspace{-2em}
\begin{center}* * *\end{center}
For our proofs, we will also use an important result from probability theory: Jensen's inequality for concave functions. We use the standard notation $\mathbb{E}(X)$ for the expected value of a random variable $X$.
\begin{theorem}[Jensen's inequality]
    Suppose $X$ is a random variable and $f:\mathbb{R}\to\mathbb{R}$ a concave function. It holds that
    \begin{equation*}
        f(\mathbb{E}(X))\geq \mathbb{E}(f(X)).
    \end{equation*}
In particular for a random variable $X$ with binomial distribution $B(n,p)$, we have
\begin{equation}\label{eq:jen_bin}
    f(np)\geq\sum_{i=0}^n \binom{n}{i}p^i(1-p)^{n-i} f(i).
\end{equation}
\end{theorem}
We will use the following result repeatedly. 
\begin{lemma}\label{Lemma:Concave}
   Let $g:\mathbb{R}_{\geq 0} \to \mathbb{R}$ given by $g:x\mapsto\sqrt{1-a^x}$. $g$ is concave on $\mathbb{R}_{\geq 0}$ if $a\in[0,1]$.
\end{lemma}
\proof
To prove the claim, consider the first derivative
\begin{align*}
    \frac{dg}{dx}(y) = \frac{\ln(a^{-1})a^y}{2\sqrt{1-a^y}}.
\end{align*}
The first derivative is non-negative, the numerator is non-increasing, and the denominator is non-decreasing in $\mathbb{R}_{\geq 0}$ since $a\in[0,1]$. Hence, the first derivative is non-increasing, which implies that $g$ is concave.\qed
%\vspace{-2em}\sectionbreak[* * *]\vspace{-2em}
\begin{center}* * *\end{center}
Note that we will consider a different but equivalent notation for the mixture of the actual quantum output of the verification protocol and the abort state. Many works denote such a state as $p \sigma + (1-p) \ketbra{\bot}$, where $p$ is the acceptance probability, and require that $\sigma$ lives in a subspace that is orthogonal to $\ketbra{\bot}$. In contrast, we make this orthogonality more explicit by writing $p \sigma \oplus (1-p)$. Hence, the state in our notation obeys a direct sum structure of the space of the quantum output and a one-dimensional abort space in which $1-p$ is a sub-normalized state.\\ 
We further shall explain the concept of an average state: The average state is the outcome of the protocol averaged over all branches (i.e., all sampling processes and measurement outcomes), where each branch is weighted with its probability. By the commutation of the sum and the direct sum, it holds that the average state for a verification protocol is given by
\begin{align}\label{eq:averageState}
    \rho_{\rm av} = \left(\sum_{e\in E}q_e p_e \rho_e\right) \oplus \sum_{e\in E}q_e(1-p_e),
\end{align}
where $E$ is the set of all branches, $q_e$ is the probability of this branch, $p_e$ is the acceptance probability and $\rho_e$ is the output for each branch.

\section{No-go results with i.i.d. attacks}\label{sec:ng-iid}
    We consider a generic protocol with $N+1$ rounds, where in each round, the source sends one register to the clients. The clients sample according to a distribution $\omega$ which register $i$ they use for the output and perform a measurement $\mu_i = \{\mu_i(0),\mathbbm{1}-\mu_i(0)\}$ on the other registers. If the outcome is $0$, the clients output the remaining register, otherwise they output the abort signal $\ketbra{\bot}$.
    \begin{algorithm}[H]
    \caption{Generic cut-and-choose protocol.}\label{proto:simple}
    \begin{algorithmic}
        \State \textbullet $\phi\in D(\X)$ is the target state,
        \State \textbullet $N+1$ is the number of rounds,
        \State \textbullet $\omega$ is a probability distribution on $\{1,...,N+1\}$,
        \State \textbullet $\{\mu_i(0)\in P(\X^{\otimes N})\}_{1\leq i\leq N+1}$ is a set of measurement operators.
    \end{algorithmic}
    \begin{algorithmic}[1]
        \State The source sends $N+1$ registers $\rho_1,...,\rho_{N+1}$.
        \State The clients sample which round is used for the output: $k\gets_{\omega}\{1,...,N+1\}$
        \State The clients measure $\rho_1,...,\rho_{k-1},\rho_{k+1},...,\rho_{N+1}$ using $\{\mu_k(0),\mathbbm{1}_{\X^{\otimes N}}-\mu_k(0)\}$, the outcome is $r\in\{0,1\}$.
        \If{$r=0$}
            \State The clients output $\rho_k$.
        \Else
            \State The clients output $\ketbra{\bot}$.
        \EndIf
    \end{algorithmic}
    \end{algorithm}
    \subsection{Stand-Alone security}
    In this section, we first present the no-go result for stand-alone security with a fidelity-based security definition. The security definition we consider is rather simple: If all parties are honest, the output of the protocol should have a high fidelity with the target state. If, however, the source is dishonest, the clients can abort, which is why we consider the optimal abort probability and compute the fidelity of the output with the target state probabilistically mixed with the abort state. As mentioned in Section \ref{sec:QITPrelim}, we expand the space $\phi$ lives in by the one-dimensional abort space using the direct sum. 
    The average states of the clients when the source is honest or dishonest, $\rho_H$ and $\rho_D$, have the same structure: The first part of the direct sum represents the actual quantum state the clients get if they accept and the second part is the probability that they reject. To simplify the notation we don't write out the implicit one dimensional abort state, and only write the abort probability.
    
    \begin{definition}\label{def:StAl_sec}
        Consider a quantum state verification protocol with a target state $\phi$. Denote with $\rho_H$ the average state of the clients if all parties are honest. We define the protocol to be $\varepsilon_H$-correct if 
        \begin{align}\label{eq:sta_def_hon}
            F(\rho_H,\phi\ \oplus\ 0)\geq 1-\varepsilon_H.
        \end{align}
        In the case where the source is dishonest, we denote with $\rho_D$  the average state of the clients. We define a quantum state verification protocol to be $\varepsilon_D$-secure against a dishonest source if 
        \begin{align}\label{eq:sta_def_dis}
            \max_{p\in[0,1]} F(\rho_D,p\phi\ \oplus\ (1-p))\geq 1-\varepsilon_D.
        \end{align}
    \end{definition}
    Intuitively, \eqref{eq:sta_def_dis} demands that the average output state for any attack the source conducts is at most $\varepsilon_D$ different from the closest state the client could get if there was the possibility to abort (with probability $1-p$) without replacing the actual quantum register. So while \eqref{eq:sta_def_hon} punishes a high abort probability if the source is honest, a high abort probability should not be disadvantageous if the source is dishonest. Further, note that the maximization over the ideal acceptance probability $p\in[0,1]$ in \eqref{eq:sta_def_dis} is only of notational relevance and does not imply a deviation from the security definition as used in e.g. \cite{colisson2024graph}. In fact, it holds that
    \begin{align*}
        \left(\max_{p\in[0,1]} F(\rho_D,p\phi\ \oplus\ (1-p))\geq 1-\varepsilon_D\right) \Leftrightarrow \left(\vphantom{\max_{p\in[0,1]}}\exists p\in[0,1]:\ F(\rho_D,p\phi\ \oplus\ (1-p))\geq 1-\varepsilon_D\right).
    \end{align*}
    And since $F(\rho_D,p\phi\ \oplus\ (1-p))$ is continuous in $p$, as one can see using \eqref{eq:fidOplus}, \eqref{eq:fidLin} and \eqref{eq:averageState}, and $[0,1]$ is a compact set, the maximum always exists. However, providing an upper bound one has to consider the maximization over $p\in[0,1]$ either way.\\[1em]

    Now we are equipped to prove one of our main results; the trade-off for cut-and-choose quantum state verification protocols with regard to stand-alone security as defined in Definition \ref{def:StAl_sec}. Theorem \ref{theorem:FidNGsimple} formalizes this trade-off, i.e., that cut-and-choose quantum state verification cannot be simultaneously correct, secure, and efficient. More specifically, \eqref{eq_tradeOffFid} states that a low incorrectness, corresponding to a low $\varepsilon_H$, implies a high insecurity $\varepsilon_D$. The efficiency of the protocol is given by the number of rounds, and the lower bound scales inversely to the round number: The fewer rounds the protocol requires, the higher the lower bound for the incorrectness and insecurity.
    \begin{theorem}\label{theorem:FidNGsimple}
        Let $\pi = (\pi_C,\pi_S)$ be a protocol as described in \ref{proto:simple}. If $\pi$ is $\varepsilon_H$-correct and $\varepsilon_D$-secure according to definition \ref{def:StAl_sec}, it holds
        \begin{align}\label{eq_tradeOffFid}
            \varepsilon_H+\varepsilon_D\geq \frac{1}{7N}.
        \end{align}
    \end{theorem}
    \proof
    In the proof, we use the property of the fidelity under the direct sum for both settings, the honest one and the dishonest one, and simplify and add the resulting equations. Finally, the i.i.d. property of the attack allows us to discard the probabilities $\omega(n)$, and known inequalities and optimized choices for the parameter of the attack yield the result.\\ 
    We start by denoting
    \begin{align*}
        p^H_A &\coloneqq \sum_{i=1}^{N+1}\omega(i)\braket{\mu_i(0)}{\phi^{\otimes N}},\\
        p^D_A &\coloneqq \sum_{i=1}^{N+1}\omega(i)\braket{\mu_i(0)}{\psi^{\otimes N}},
    \end{align*}
    where $\psi$ is the state of which a dishonest source sends $N+1$ copies in an i.i.d. attack. Then
    $p^H_A$ and $p^D_A$ are the average probabilities that the clients accept the verification in the honest and dishonest case, respectively.
    We find
    \begin{align*}
        \rho_H = p^H_A \phi\ \oplus\ (1-p^H_A),
    \end{align*}
    which implies by \eqref{eq:fidOplus}
    \begin{align}\label{eq:FidHonest}
        F(\rho_H,\phi\oplus 0) = p^H_A\Rightarrow \varepsilon_H\geq 1-p^H_A.
    \end{align}
    If the source is dishonest and sends $N+1$ copies of $\psi$ we find
    \begin{align*}
        \rho_D = p^D_A \psi\ \oplus\ (1-p^D_A),
    \end{align*}
    which implies again by \eqref{eq:fidOplus}
    \begin{align*}
        &\max_{p\in[0,1]}F(\rho_D,p\phi\ \oplus\ (1-p)) =\max_{p\in[0,1]}\left(\sqrt{p^D_Ap}\sqrt{F(\phi,\psi)}+\sqrt{(1-p^D_A)(1-p)}\right)^2\\
        \Rightarrow &\varepsilon_D\geq \min_{p\in[0,1]}\left(1-\left(\sqrt{p^D_Ap}\sqrt{F(\phi,\psi)}+\sqrt{(1-p^D_A)(1-p)}\right)^2\right).
    \end{align*}
    When considering 
    \begin{align*}
        f(p) \coloneqq \sqrt{p}a+\sqrt{1-p}b
    \end{align*}
    we find for $a,b\geq 0$ and $0 < p' < 1$:
    \begin{align*}
        &\frac{d}{dp}f(p') = \frac{a}{2\sqrt{p'}}-\frac{b}{2\sqrt{1-p'}}=0\Leftrightarrow \frac{a^2}{p'} = \frac{b^2}{1-p'} \Leftrightarrow p' = \frac{a^2}{a^2+b^2}.
    \end{align*}
    Using this for maximizing $f^2$ yields
    \begin{align*}
        \max_{p\in[0,1]} f(p)^2 = \left(\frac{a^2}{\sqrt{a^2+b^2}} + \frac{b^2}{\sqrt{a^2+b^2}}\right)^2 = a^2+b^2.
    \end{align*}
    Using $a=\sqrt{p^D_AF(\phi,\psi)}$ and $b=\sqrt{1-p^D_A}$
    gives
    \begin{align*}
        \varepsilon_D &\geq 1-\max_{p\in[0,1]}F(\rho_D,p\phi\ \oplus\ (1-p)) = 1-\left(p^D_AF(\phi,\psi) + 1-p^D_A\right) = p^D_A(1-F(\phi,\psi)),
    \end{align*}
    which is smaller than $1-F(\rho_D,\phi\ \oplus\ 0) = 1-p^D_AF(\phi,\psi)$ and $1-F(\rho_D,(0\cdot\phi)\ \oplus\ 1) = p^D_A$.
    Combined with the bound for $\varepsilon_H$, we find that
    \begin{align*}
        \varepsilon_H+\varepsilon_D&\geq 1-p^H_A+p^D_A(1-F(\phi,\psi)) \geq (1-F(\phi,\psi))\left(1-p^H_A\right) + (1-F(\phi,\psi))p_A^D\\ &= (1-F(\phi,\psi))\left(1-\left(p^H_A-p^D_A\right)\right) \geq (1-F(\phi,\psi))\left(1-\left|p^H_A-p^D_A\right|\right).
    \end{align*}
    $|p^H_A-p^D_A|$ is the distinguishing advantage, i.e. the absolute value of the difference of the measurement probabilities, when distinguishing $\phi^{\otimes N}$ and $\psi^{\otimes N}$ using $\mu_i$ averaged over $i$. Therefore, using the Holevo-Helstrom theorem (in particular (\ref{eq:dist_adv})) and \eqref{eq:Multi_copy_dist} we find
       \begin{align}\label{eq:pdiffSimp}
        |p^H_A-p^D_A| &\leq \sum_{i=1}^{N+1}\omega(i) \left|\braket{\mu_i(0)}{\phi^{\otimes N}}-\braket{\mu_i(0)}{\psi^{\otimes N}}\right|\nnn &\leq \frac{1}{2}\left\|\phi^{\otimes N}-\psi^{\otimes N}\right\|\sum_{i=1}^{N+1}\omega(i)\leq \sqrt{1-F(\phi,\psi)^N}.
    \end{align}
    Combined with the above, this means
    \begin{align*}
        \varepsilon_H+\varepsilon_D\geq (1-F(\phi,\psi))\left(1-\sqrt{1-F(\phi,\psi)^N}\right).
    \end{align*}
    Now we define $\tau \coloneqq (1-F(\phi,\psi))$, i.e. $F(\phi,\psi) = (1-\tau)$:
    \begin{align*}
        \varepsilon_H+\varepsilon_D\geq \tau\left(1-\sqrt{1-(1-\tau)^{N}}
        \right).
    \end{align*}
    With $\alpha \in [0,1]$, we choose $\tau = \nicefrac{\alpha}{N}$, and using $(1-\nicefrac{\alpha}{N})^N\geq 1-\alpha$  we get
    \begin{align}\label{eq:fidSimPreBound}
        \varepsilon_H+\varepsilon_D\geq \frac{\alpha}{N}\left(1-\sqrt{1-\left(1-\frac{\alpha}{N}\right)^N}\right)\geq \frac{\alpha}{N}\left(1-\sqrt{\alpha}\right)\eqqcolon h_N(\alpha),
    \end{align}
    which is maximized for $\alpha=\nicefrac{4}{9}$ since
    \begin{align*}
        \frac{dh_N}{d\alpha} (\alpha_0) = N-\frac{3N}{2}\sqrt{\alpha_0} = 0 \Rightarrow \alpha_0 = \frac{4}{9}
    \end{align*} and $h_N(0)=h_N(1)=0$.\\
    This in turn yields
    \begin{align*}
        \varepsilon_H+\varepsilon_D\geq \frac{4}{27N} \geq \frac{1}{7N}.
    \end{align*}\qed
    \subsection{Composable security}
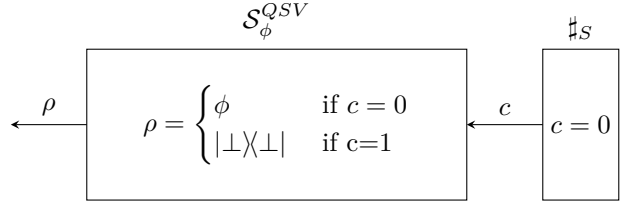
\begin{figure}[H]
        \centering
        \begin{tikzpicture}
        \draw (0,0) rectangle (5,2) node[pos=0.5, align=left] {$\rho = \begin{cases}\phi&\text{ if } c=0\\
            \ketbra{\bot} &\text{ if c=1}\end{cases}$};
        \draw[-{stealth}] (0,1) -- (-1,1) node[pos=0.5,above] {$\rho$};
        \draw[{stealth}-] (5,1) -- (6,1) node[pos=0.5,above] {$c$};
        \node[anchor=south] at (2.5,2) {$\S_{\phi}^{QSV}$};
        \draw (6,0) rectangle (7,2) node[pos=0.5] {$c=0$};
        \node[anchor=south] at (6.5,2) {$\sharp_S$};
    \end{tikzpicture}
        \caption{Ideal quantum state verification: The source chooses whether to be honest or not by inputting a bit $c\in\{0,1\}$ to the ideal resource $\S_{\phi}^{QSV}$. If it is honest, $c$ is $0$ and the ideal resource sends the target state $\phi$ to the clients. If the source is dishonest, $c$ can be set to $1$ and the clients receive an abort signal. The right box $\sharp_S$ represents a filter that enforces honest behavior of the ideal resource if the source is honest, i.e. $\sharp_S$ fixes the input $c=0$. In case the source is dishonest, we ignore the filter which allows the source to input $c=1$.}
        \label{fig:ideal}
    \end{figure}

    We consider composable security definitions following the `real world vs. ideal world' paradigm. Such security definitions have been presented for \emph{universal composability} (UC) \cite{Canetti}, \emph{abstract cryptography} (AC) \cite{MauRen11}, and \emph{categorical composable cryptography} \cite{broadbentkarvonen:categoricalcrypto,broadbentkarvonen:categoricalcryptoextended} (CCC); for every attack on the implementation, there has to be an attack on the ideal resource which makes the two settings indistinguishable up to some $\varepsilon\geq 0$. While in AC and UC, a simulator translates attacks on the implementation into attacks on the ideal resource, in CCC, the user chooses the ideal attack more freely. We omit the actual security definitions and focus on finding a lower bound on the trace distance between ideal and real output states, which translates into a no-go result for the above-mentioned frameworks for composable cryptography.
    Agnostic of the actual framework, we consider the ideal resource in Fig. \ref{fig:ideal} for quantum state verification for a target state $\phi$.
    Regarding correctness, the Holevo-Helstrom theorem (Theorem \ref{thm:holevo_helstrom}) states that the distinguishing advantage when given $\phi$ from the ideal resource or $\rho_H$ from the implementation, is given by the trace distance, i.e. $\varepsilon_H$-correctness implies
    \begin{align*}
        \frac{1}{2}\left\|\rho_H-(\phi\oplus 0)\right\|_1\leq \varepsilon_H.
    \end{align*}
    For security, if the source is dishonest, we need to consider all possible ideal attacks or simulators, respectively. However, all an ideal attack or simulator can do at the interface of the ideal resource is to input $c=0$ with some probability $p$ which might depend on the attack on the implementation. While a distinguisher (or attacker) has the freedom to use a private register to increase the distinguishing advantage, it suffices for our no-go result to omit such register, i.e. composable security up to $\varepsilon_D$ implies with $\rho_D$ being the output state of the implementation
    \begin{align*}
        \min_{p\in[0,1]}\frac{1}{2}\left\|\rho_D-(p\phi\oplus (1-p)\right\|_1\leq \varepsilon_D
    \end{align*}
    but not necessarily vice-versa. Note that the minimization over all probabilities comes from the choice of the ideal attack or simulator, respectively. 
    \begin{theorem}\label{theorem:composSimple}
    Let $\pi = (\pi_C,\pi_S)$ be a protocol as described in Protocol \ref{proto:simple}. If 
    \begin{align*}
        \frac{1}{2}\left\|\rho_H-(\phi\oplus 0)\right\|_1\leq \varepsilon_H
    \end{align*}
    and
    \begin{align*}
        \min_{p\in[0,1]}\frac{1}{2}\left\|\rho_D-(p\phi\oplus (1-p)\right\|_1\leq \varepsilon_D
    \end{align*}
    It holds $\varepsilon_H+\varepsilon_D\geq \frac{\sqrt{\eta_1}}{4\sqrt{N}}$ if $\phi$ is mixed and $\eta_1$ is the largest eigenvalue of $\phi$ and $\varepsilon_H+\varepsilon_D\geq \frac{1}{4\sqrt{N}}$ if is $\phi$ pure.
    \end{theorem}
    \proof
        Similar to Theorem \ref{theorem:FidNGsimple}, we use the properties of the trace distance for each setting, honest and dishonest source, separately, and add the resulting lower bounds. Next, we fix that type of states that the source uses for the attack and derive in an intricate analysis, a lower bound for the trace distance of $N$ copies of the target state and the state the dishonest source sends, which is similar to the trace distance of $N$ pure states. Finally, known inequalities and optimized parameter choices prove the lower bound claimed in the theorem.\\
        We first note that for any positive semidefinite operators $P_0,P_1,Q_0$ and $Q_1$ it holds
        \begin{align*}
            \left\|(P_0\oplus P_1)-(Q_0\oplus Q_1)\right\|_1 = \left\|P_0-Q_0\right\|_1 +\left\|P_1-Q_1\right\|_1.
        \end{align*}
        As before, we denote 
        \begin{align*}
        p^H_A &\coloneqq \sum_{i=1}^{N+1}\omega(i)\braket{\mu_i(0)}{\phi^{\otimes N}}\\
        p^D_A &\coloneqq \sum_{i=1}^{N+1}\omega(i)\braket{\mu_i(0)}{\psi^{\otimes N}}.
    \end{align*}
    We find
    \begin{align*}
        \rho_H = p^H_A \phi\ \oplus\ (1-p^H_A),
    \end{align*}
    which implies
    \begin{align*}
       \varepsilon_H\geq  \frac{1}{2}\|\rho_H-(\phi\oplus 0)\| = 1-p^H_A.
    \end{align*}
    If the source is dishonest and sends $N+1$ copies of $\psi$ we find
    \begin{align*}
        \rho_D = p^D_A \psi\ \oplus\ (1-p^D_A),
    \end{align*}
    which implies, using the triangle inequality
    \begin{align*}
    \varepsilon_D\geq&\min_{p\in[0,1]}\frac{1}{2}\| \rho_D - (p\phi\ \oplus\ (1-p))\|_1 = \min_{p\in[0,1]} \frac{1}{2}\left\|p^D_A\psi - p\phi\right\|_1 + \frac{1}{2}|p-p^D_A|\\ 
        =&\min_{p\in[0,1]} \frac{1}{2}\left\|p^D_A\psi - p\phi\right\|_1 + \frac{1}{2}\|p\phi-p^D_A\phi\|_1\geq  \frac{p^D_A}{2}\left\|\psi-\phi\right\|_1.
    \end{align*}
    Now, we can combine both settings and find
    \begin{align*}
        \varepsilon_H+\varepsilon_D&\geq \frac{p^D_A}{2} \left\|\psi-\phi\right\|_1 + 1 - p^H_A\geq \frac{1}{2}\left\|\psi-\phi\right\|_1\left(1-\left|p^H_A-p^D_A\right|\right).
    \end{align*}
    Using \eqref{eq:pdiffSimp}, we find
    \begin{align}\label{eq:combined}
        \varepsilon_H+\varepsilon_D&\geq \frac{1}{2}\left\|\psi-\phi\right\|_1\left(1-\frac{1}{2}\left\|\phi^{\otimes N} - \psi^{\otimes N}\right\|_1\right)
    \end{align}
    In the next step, we fix $\psi$. We write $\phi$ in spectral decomposition as $\phi=\sum_{i=1}^d\eta_i\ketbra{\phi_i}$ with $d=\rank(\phi)$ and $\eta_i\geq \eta_{i+1}$ for $1\leq i<d$. For some pure state $\ket{\chi}$, we fix $\psi = \eta_1 \ketbra{\chi} + \sum_{i=2}^d\eta_i\ketbra{\phi_i}$. This implies 
    \begin{align}
        \frac{1}{2}\left\|\phi-\psi\right\|_1 = \eta_1\sqrt{1-|\braket{\phi_1}{\chi}|^2}.
    \end{align}
    In the case of $N$ copies, things become more complicated. We first note that
    \begin{align*}
        \phi^{\otimes N} = \sum_{A\in\{1,...,d\}^N} \bigotimes_{i\in A}\eta_i\ketbra{\phi_i},
    \end{align*}
    and similarly with $\ketbra{\chi_i}\coloneqq\ketbra{\phi_i}$ for $i\neq 1$ and $\ketbra{\chi_1}\coloneqq\ketbra{\chi}$
    \begin{align*}
        \psi^{\otimes N} = \sum_{A\in\{1,...,d\}^N} \bigotimes_{i\in A}\eta_i\ketbra{\chi_i}.
    \end{align*}
    Hence, we find
    \begin{align*}
        \phi^{\otimes N}-\psi^{\otimes N} = \sum_{A\in\{1,...,d\}^N}\left[\bigotimes_{i\in A}\eta_i\ketbra{\phi_i}-\bigotimes_{i\in A}\eta_i\ketbra{\chi_i}\right].
    \end{align*}
     Now consider the trace norm of this difference. We find an upper bound using the triangle inequality in a way, such that we group the terms for which the positions of the $1$s in $A$ are the same, i.e. we get $2^N$ trace differences. Next, using the unitary invariance of the trace norm, we can move all positions corresponding to $1$ in $A$ to the front. Next, we use 
    \begin{align*}
        \|\rho_1^{\otimes i} \otimes \nu^{\otimes (N-i)} - \rho_2^{\otimes i} \otimes \nu^{\otimes (N-i)}\|_1 = \|\rho_1^{\otimes i}-\rho_2^{\otimes i}\|_1\|\nu^{\otimes (N-i)}\|_1, 
    \end{align*}
    with $\rho_1 = \eta_1\ketbra{\phi_1},\ \rho_2 = \eta_1\ketbra{\chi_1}$ and $\nu = \sum_{i=2}^N\eta_i\ketbra{\phi_i}$. Now, we note that $\|\nu^{\otimes (N-i)}\|_1 = (1-\eta_1)^{N-i}$. So we find
    \begin{align*}
        &\frac{1}{2}\left\|\phi^{\otimes N}-\psi^{\otimes N}\right\|_1\leq \sum_{i=0}^N\binom{N}{i}\frac{1}{2}\left\|\ketbra{\phi_1}^{\otimes i}-\ketbra{\chi}^{\otimes i}\right\|_1\eta_1^i(1-\eta_1)^{N-i}.
    \end{align*}
    Using the expression for the trace distance of pure states \eqref{eq:trace_pure} and Jensen's inequality for concave functions with a binomial distribution \eqref{eq:jen_bin}, with $f:x\to \sqrt{1-|\braket{\phi_1}{\chi}|^{2x}}$ (cf. Lemma \ref{Lemma:Concave}) and binomial distribution $B(N,\eta_1)$, we find
    \begin{align*}
        \frac{1}{2}\left\|\phi^{\otimes N}-\psi^{\otimes N}\right\|_1 &\leq \sum_{i=0}^N\binom{N}{i}\sqrt{1-|\braket{\phi_1}{\chi}|^{2i}}\eta_1^i(1-\eta_1)^{N-i}\\
        &\leq \sqrt{1-|\braket{\phi_1}{\chi}|^{2\eta_1N}}.
    \end{align*}
    So for $\varepsilon_H+\varepsilon_D$ we find
    \begin{align*}
        \varepsilon_H+\varepsilon_D\geq \eta_1\sqrt{1-|\braket{\phi_1}{\chi}|^2}\left(1-\sqrt{1-|\braket{\phi_1}{\chi}|^{2\eta_1N}}\right).
    \end{align*}
    Writing $\tau\coloneqq\sqrt{1-|\braket{\phi_1}{\chi}|^2}$ yields
    \begin{align*}
        \varepsilon_H+\varepsilon_D\geq \eta_1\tau\left(1-\sqrt{1-(1-\tau^2)^{\eta_1N}}\right).
    \end{align*}
    Choosing $\tau=\nicefrac{\alpha}{\sqrt{\eta_1 N}}$ gives us using $(1-a/N)^N\geq 1-a$
    \begin{align*}
        \varepsilon_H+\varepsilon_D\geq \nicefrac{\alpha\sqrt{\eta_1}}{\sqrt{N}}\left(1-\alpha\right).
    \end{align*}
    The optimal value for $\alpha$ is $\nicefrac{1}{2}$, hence we find
    \begin{align*}
        \varepsilon_H+\varepsilon_D\geq\frac{\sqrt{\eta_1}}{4\sqrt{N}}.
    \end{align*}
    In particular, for pure states this gives us
    \begin{align*}
        \varepsilon_H+\varepsilon_D\geq\frac{1}{4\sqrt{N}}.
    \end{align*}
    \qed
    
    This provides directly a lower bound for composable security definitions as used in \cite{MauRen11,broadbentkarvonen:categoricalcrypto}. The distinguisher can implement the presented attack without an auxiliary register. By design of the ideal resource, the simulator can only accept or reject with some probability; hence, the relevant measure for this attack is in fact the trace distance.

   \section{Optimal attacks}\label{sec:ng-non-iid}
        With a similar proof technique, we now prove our no-go result for more complex attacks without the i.i.d. restriction. We study optimal attacks, attacks leading to the highest security violation, i.e. that maximize $\varepsilon_H+\varepsilon_D$. By dropping the i.i.d. restriction, we study the case where the dishonest source is free to send an arbitrary state in every round. Nevertheless, we still consider the attacker, i.e. the dishonest source, to act in the class of separable state attacks. More precisely, we demand that there is no entanglement between states from different rounds, i.e., an attacker sends separable states $\{\psi_j\}_{j=1}^{N+1}$. 
        This class, in fact, contains a naive attack that could lead to the clients sharing a state that is orthogonal to the desired output, which is the maximal deviation possible. In this naive attack the source guesses which round will be used for the output and sends for this particular round an orthogonal state, but for all other rounds it sends the actual target state to minimize the risk of getting caught.
        
        This naive attack and its comparison to the i.i.d. attacks motivate this investigation.
        \begin{lemma}\label{lemma:equal}
            If $\phi$ in Protocol \ref{proto:simple} is pure and an attacker sends separable states $\{\psi_j\}_{j=1}^{N+1}$, it holds
            \begin{align*}
                F(\rho_H,\phi) &= \sum_{i=1}^{N+1}\omega(i)p_H(i)\\
                \max_{p\in[0,1]}F(\rho_D,p\phi\oplus (1-p)) &= 1-\sum_{i=1}^{N+1}\omega(i)p_D(i)(1-F(\psi_i,\phi)),
            \end{align*}
            where $\rho_H$ is the output state in the honest setting, $\rho_D$ the one if the source cheats and
            \begin{align*}
                p_H(i)&\coloneqq\braket{\mu_i(0)}{\bigotimes_{j=1,j\neq i}^{N+1}\phi}\\
                p_D(i)&\coloneqq\braket{\mu_i(0)}{\bigotimes_{j=1,j\neq i}^{N+1}\psi_j}.
            \end{align*}
        \end{lemma}
        \proof The proof uses the property of the fidelity under the direct sum and results which we introduced in Theorem \ref{theorem:FidNGsimple}.\\
        In Protocol \ref{proto:simple}, we find if the source is honest:
        \begin{align*}
            \rho_H = \left(\sum_{i=1}^{N+1}\omega(i) p_H(i)\right)\phi \oplus \left(1-\sum_{i=1}^{N+1} \omega(i) p_H(i)\right),  
        \end{align*}
        and
        \begin{align*}
            F(\rho_H,\phi\oplus 0)
            = \sum_{i=1}^{N+1}\omega(i)p_H(i),
        \end{align*}
        which is already the first part of the statement.
        In the dishonest setting, we find
        \begin{align*}
            \rho_D = \sum_{i=1}^{N+1} \omega(i) p_D(i) \psi_i \oplus \left(1-\sum_{i=1}^{N+1}\omega(i)p_D(i)\right). 
        \end{align*}
        Similarly, as in the previous section, this implies
        \begin{align*}
            &\max_{p\in[0,1]}F(\rho_D,p\phi\oplus 1-p) = \left(\sqrt{p}\sqrt{F\left(\sum_{i=1}^{N+1}\omega(i)p_D(i)\psi_i,\phi\right)} + \sqrt{1-p}\sqrt{1-\sum_{i=1}^{N+1}\omega(i)p_D(i)}\right)^2.
        \end{align*}
        Using $\max_{p\in[0,1]}(\sqrt{p}a+\sqrt{1-p}b)^2 = a^2+b^2$, we find
        \begin{align*}
            \max_{p\in[0,1]}F(\rho_D,p\phi\oplus 1-p) = F&\left(\sum_{i=1}^{N+1}\omega(i)p_D(i)\psi_i,\phi\right)+\left(1-\sum_{i=1}^{N+1}\omega(i)p_D(i)\right).
        \end{align*}
        The assumption that $\phi$ is pure implies  
        \begin{align*}
            &\max_{p\in[0,1]}F(\rho_D,p\phi\oplus 1-p) = \sum_{i=1}^{N+1}\omega(i)p_D(i)\bra{\phi}\psi_i\ket{\phi} + \left(1-\sum_{i=1}^{N+1}\omega(i)p_D(i)\right)\\ &= \sum_{i=1}^{N+1}\omega(i)p_D(i)F(\psi_i,\phi) + \left(1-\sum_{i=1}^{N+1}\omega(i)p_D(i)\right)= 1-\sum_{i=1}^{N+1}\omega(i)p_D(i)(1-F(\psi_i,\phi)).
        \end{align*}\qed\\[1em]

        \begin{lemma}\label{lemma:genlowbound}
            Let $\pi$ be a quantum state verification protocol as described in Protocol \ref{proto:simple} for a pure target state $\phi$, $\varepsilon_H$ be the lowest value that fulfills \eqref{eq:sta_def_hon} and $\varepsilon_D$ be the lowest value that fulfills \eqref{eq:sta_def_dis}, i.e., the best possible security parameter in fidelity-based security (Definition \ref{def:StAl_sec}).
            If an attacker sends separable states $\{\psi_j\}_{j=1}^{N+1}$, it holds
            \begin{align}\label{eq:gen_low_bound}
            \varepsilon_H+\varepsilon_D\geq \sum_{i=1}^{N+1}\omega(i) (1-F(\psi_i,\phi)) \left(1-\sqrt{1-\prod_{j=1,j\neq i}^{N+1}F(\phi,\psi_j)}\right),
        \end{align}
        where the inequality is saturated if the protocol uses optimal measurements to distinguish the target state from any other state, and the source sends pure states.
        \end{lemma}
        \proof The result follows from the previous lemma, the Fuchs-van de Graaf inequalities, and the Holevo Helstrom theorem.
        We start by using Lemma \ref{lemma:equal} and find find 
        \begin{align*}
            \varepsilon_H+\varepsilon_D &= \sum_{i=1}^{N+1}\omega(i)\left(1-p_H(i)\right) + \sum_{i=1}^{N+1}\omega(i)p_D(i)(1-F(\psi_i,\phi)) \\
            &= \sum_{i=1}^{N+1}\omega(i) \left(\left(1-p_H(i)\right) + p_D(i)(1-F(\psi_i,\phi))\right).
        \end{align*}
        If the protocol uses optimal measurements to distinguish a pure target state from any other state, then $p_H(i)=1$, which implies $(1-F(\psi_i,\phi))\left(1-p_H(i)\right) = 1-p_H(i)$. If the protocol uses other measurements it holds that $(1-F(\psi_i,\phi))\left(1-p_H(i)\right) \leq 1-p_H(i)$. Hence,
        \begin{align*}
            \varepsilon_H+\varepsilon_D &= \sum_{i=1}^{N+1}\omega(i)\left(1-p_H(i)\right) + \sum_{i=1}^{N+1}\omega(i)p_D(i)(1-F(\psi_i,\phi)) \\
            &\geq \sum_{i=1}^{N+1}\omega(i) (1-F(\psi_i,\phi))\left(1-\left(p_H(i) - p_D(i)\right)\right),
        \end{align*}
        where the inequality is saturated if the protocol uses optimal measurements. One finds by \eqref{eq:holevo_helstrom}, \eqref{eq:Multi_copy_dist} and Theorem \ref{thm:FuchsVDG} that
        \begin{align*}
            p_H(i) - p_D(i)\leq |p_H(i) - p_D(i)| \leq \frac{1}{2}\left\|\bigotimes_{j=1,j\neq i}^{N+1}\psi_j - \bigotimes_{j=1,j\neq i}^{N+1}\phi\right\|_1\leq \sqrt{1-\prod_{j=1,j\neq i}F(\phi,\psi_j)},
        \end{align*} where the inequalities are saturated if the measurements are optimal, and the states the source sent are pure. Hence, one finds that
        \begin{align*}
            \varepsilon_H+\varepsilon_D &\geq \sum_{i=1}^{N+1}\omega(i) (1-F(\psi_i,\phi))\left(1-\sqrt{1-\prod_{j=1,j\neq i}F(\phi,\psi_j)}\right), 
        \end{align*}
        where under the mentioned assumptions the inequality is saturated.
        \qed\\[1em]
        
        We show that an attack in which the attacker sends an orthogonal state in one round and the target state in all other rounds maximizes this bound.
        \begin{theorem}\label{theorem:optimalFidAttack} 
            Assume $\phi$ is pure, hence, there exists at least one state $\phi^{\perp}$ such that $F(\phi,\phi^{\perp})=0$.
            This allows for an attack where $F(\psi_j,\phi)=1-\delta_{j,\ell}$ with $\max_{1\leq j\leq N+1}\omega(j) = \omega(\ell)$. 
             This attack is optimal in the class of separable state attacks with regard to the bound presented in \eqref{eq:gen_low_bound}.
        \end{theorem}
        \proof We first derive an inequality which implies that the bound presented in \eqref{eq:gen_low_bound} cannot exceed $\omega_{\ell}$. We prove this inequality using the non-negativity and concavity of $\log(x+1)$ and eventually demonstrate that the attack described in the claim yields exactly $\omega_{\ell}$ inserted in the bound presented in \eqref{eq:gen_low_bound}.
        
        We denote the right hand side of~\eqref{eq:gen_low_bound} as the function $B_{N}:[0,1]^{N+1}\to [0,1]$ and first prove that for $\mathbf{f}\in (0,1]^{N+1}$ we have
        \begin{align}\label{eq:boundForbound}
            B_{N}(\mathbf{f}) \coloneqq \sum_{i=1}^{N+1}\omega(i) (1-f_i) \left(1-\sqrt{1-\prod_{j=1,j\neq i}^{N+1}f_j}\right)\leq \omega(\ell).
        \end{align}
        We use $1-\sqrt{1-\prod_{j=1,j\neq i}^{N+1}f_j}\leq \prod_{j=1,j\neq i}^{N+1}f_j$ and find
        \begin{align*}
            B_{N}(\mathbf{f}) &\leq \sum_{i=1}^{N+1} \omega(i)(1-f_i) \left(\prod_{j=1,j\neq i}^{N+1}f_j\right) = \left(\prod_{j=1}^{N+1}f_j\right)\sum_{i=1}^{N+1}\omega(i)\frac{1-f_i}{f_i} \leq \omega(\ell) \left(\prod_{j=1}^{N+1}f_j\right)\sum_{i=1}^{N+1}\frac{1-f_i}{f_i}.
        \end{align*}
        So \eqref{eq:boundForbound} follows from
        \begin{align}\label{eq:almostThere}
            \sum_{i=1}^{N+1}\frac{1-f_i}{f_i} = \sum_{i=1}^{N+1}\left(\frac{1}{f_i}-1\right)\leq\prod_{i=1}^{N+1}\frac{1}{f_i},
        \end{align}
        
    which we now prove. First note that $\log(x+1)$ is non-negative and is concave, i.e. for $a, b\geq 0$
        \begin{align*}
            \log(a+b+1) \leq \log(a+1)+\log(b+1),
        \end{align*}
            meaning $\log(x+1)$ is sub-additive. Further, $\log$ is strictly increasing and thus preserves inequalities. Hence, we find for (\ref{eq:almostThere})
        \begin{align*}
            \sum_{i=1}^{N+1}\left(\frac{1}{f_i}-1\right)&\leq\prod_{i=1}^{N+1}\frac{1}{f_i}\Leftarrow\  \log(1+\sum_{i=1}^{N+1}\left(\frac{1}{f_i}-1\right))\leq \sum_{i=1}^{N+1}\log(\frac{1}{f_i}) \\&= \log(\prod_{i=1}^{N+1}\frac{1}{f_i})\leq\log(1+\prod_{i=1}^{N+1}\frac{1}{f_i}),
        \end{align*}
which implies \eqref{eq:boundForbound}. 

We now consider $\mathbf{f}\in [0,1]^{N+1}$ and define the set $A_{\mathbf{f}} = \{i\mid 1\leq i\leq N+1, f_i=0\}$. We note that $B_{N}(\mathbf{f})$ is non-zero only when $A$ is empty or a singleton. Indeed, if two or more $f_k=0$ it holds that $1-\sqrt{1-\prod_{j=1,j\neq i}^{N+1}f_j} = 0$ for all $i\in\{1,...,N+1\}$. However, if $A_{\mathbf{f}}$ is a singleton, i.e. $A_{\mathbf{f}} = \{k\}$ we find $B_{N}(\mathbf{f}) = \omega(k) \left(1-\sqrt{1-\prod_{j=1,j\neq k}^{N+1}f_j}\right)$. The case where $A_f$ is empty is covered in the first half of the proof. Hence, we find that $B_{N}(\mathbf{f})\leq \omega(\ell)$ and get equality for the described attack.\qed\\[1em]
        The described attack in Theorem \ref{theorem:optimalFidAttack} maximizes the bound proved in Lemma \ref{lemma:genlowbound}. This Lemma also proves that if the protocol uses optimal measurements, the target state is pure and the source sends pure states, the bound is saturated. Hence, for such an optimal protocol for a pure target state, the presented attack is optimal in the class of separable state attacks, i.e., no other separable state attack can achieve higher violation of security. Indeed, Lemma \ref{lemma:secOfProto} in the appendix even implies that a larger security violation is not always possible, since it proves that a specific choice of parameters in Protocol \ref{proto:simple} yields $0$-correctness and $\nicefrac{1}{N+1}$-security for stand-alone security where $\omega(i) = \nicefrac{1}{N+1}$ for all $0<i\leq N+1$ was chosen. 
        While the presented attack is the intuitive and naive approach to break verification, it was unknown if it was optimal in the class of separable pure-state attacks. However, while this all holds with respect to the fidelity-based security definition, interestingly, the statement is false for composable security definitions, as we show in the remainder of this section. We start with the composable version of Lemma \ref{lemma:equal}.
        \begin{lemma}\label{lemma:composableEqual}
            If $\phi$ in Protocol \ref{proto:simple} is pure and an attacker sends separable states $\{\psi_j\}_{j=1}^{N+1}$, it holds
            \begin{align*}
                \frac{1}{2}\|\rho_H-\phi\oplus 0\|_1 &= 1-\sum_{i=1}^{N+1}\omega(i)p_H(i),\\
                \frac{1}{2}\min_{p\in[0,1]}\|\rho_D-p\phi\oplus (1-p)\|_1 &= \frac{1}{2}\left\|\sum_{i=1}^{N+1}\omega(i)p_D(i)(\psi_i-\phi)\right\|_1,
            \end{align*}
            where $\rho_H$ is the output state in the honest setting, $\rho_D$ the one if the source cheats and $p_H(i)$ and $p_D(i)$ are defined as in Lemma \ref{lemma:equal}.
        \end{lemma}
        \proof We start with the honest setting. When the source is honest, the output state is 
        \begin{align*}
            \rho_H = \sum_{i=1}^{N+1}\omega(i)p_H(i) \phi \oplus \left(1-\sum_{i=1}^{N+1}\omega(i)p_H(i)\right).
        \end{align*}
        This implies
        \begin{align*}
            \frac{1}{2}\|\rho_H-\phi\oplus 0\|_1 = \frac{1}{2}&\left\|\sum_{i=1}^{N+1}\omega(i)p_H(i) \phi - \phi\right\|_1 + \frac{1}{2}\left|1-\sum_{i=1}^{N+1}\omega(i)p_H(i)\right|= 1-\sum_{i=1}^{N+1}\omega(i)p_H(i).
        \end{align*}
        If the source is dishonest, the output state of the protocol is
        \begin{align*}
            \rho_D = \sum_{i=1}^{N+1}\omega(i)p_D(i) \psi_i \oplus \left(1-\sum_{i=1}^{N+1}\omega(i)p_D(i)\right),
        \end{align*}
        which implies 
        \begin{align*}
           & \min_{p\in[0,1]}\frac{1}{2}\|\rho_D-p\phi\oplus (1-p)\|_1 =\min_{p\in[0,1]} \frac{1}{2}\left\|\sum_{i=1}^{N+1}\omega(i)p_D(i) \psi_i - p\phi\right\|_1 + \frac{1}{2}\left|p-\sum_{i=1}^{N+1}\omega(i)p_D(i)\right|.
        \end{align*}
        On the one hand, with $p= \sum_{i=1}^{N+1}\omega(i)p_D(i)$ we find
        \begin{align*}
           \min_{p\in[0,1]} \frac{1}{2}\|\rho_D-p\phi\oplus (1-p)\|_1 
           \leq \frac{1}{2}\left\|\sum_{i=1}^{N+1}\omega(i)p_D(i)(\psi_i-\phi)\right\|_1.
        \end{align*}
        On the other hand, we find using the triangle inequality and multiplication by $1=\|\phi\|_1$:
        \begin{align*}
            &\min_{p\in[0,1]}\frac{1}{2}\|\rho_D-p\phi\oplus (1-p)\|_1 = \min_{p\in[0,1]} \frac{1}{2}\left\|\sum_{i=1}^{N+1}\omega(i)p_D(i) \psi_i - \phi\right\|_1 + \frac{1}{2}\left\|p\phi-\sum_{i=1}^{N+1}\omega(i)p_D(i)\phi\right\|_1\\ \geq &\frac{1}{2}\left\|\sum_{i=1}^{N+1}\omega(i)p_D(i)(\psi_i-\phi)\right\|_1,
        \end{align*}
        which proves the claim.\qed\\[1em]
        Now we can easily show that the naive approach of sending an orthogonal state for the round with the highest output probability and the target state in all other rounds is not always optimal. If the target state is pure and the protocol uses optimal measurements, we find for this specific attack that
        \begin{align*}
          \min_{p\in[0,1]}   \frac{1}{2}\|\rho_D-p\phi\oplus (1-p)\|_1 &=\frac{1}{2}\left\|\sum_{i=1}^{N+1}\omega(i)p_D(i)(\psi_i-\phi)\right\|_1 =\frac{\omega(\ell)}{2}\|\psi_{\ell}-\phi\|_1= \omega(\ell),\\
          \frac{1}{2}\|\rho_H-\phi\oplus 0\|_1 &= 0
        \end{align*}
        since for all $i\neq\ell$ it holds $\psi_i=\phi$ which implies  $p_D(\ell)=1$ for optimal measurements and $\psi_{\ell} = \phi^{\perp}$ which implies $\nicefrac{1}{2}\|\psi_{\ell}-\phi\|_1=1$. With the specifications of the protocol being $\omega(i)=\nicefrac{1}{N+1}$, we find for this specific attack on this protocol $\varepsilon_H+\varepsilon_D = \nicefrac{1}{N+1}<\nicefrac{1}{N}$. However, we already know that there is an i.i.d. attack such that $\varepsilon_H+\varepsilon_D\geq \nicefrac{1}{4\sqrt{N}}$ if $\phi$ is pure by virtue of Theorem \ref{theorem:composSimple}; hence, we find that the naive approach is not optimal if $N>16$. 
        
        \section{Discussion}\label{sec:disc}
        Quantum state verification is of utmost importance for quantum cryptography. We demonstrate that implementations using the popular cut-and-choose approach cannot succeed in a desirable parameter regime even if the attacker is restricted to i.i.d. attacks. Further, we find that the naive approach for protocols with a fixed number of rounds is optimal for the fidelity-based security definition but exhibits a suboptimal scaling for composable security. Indeed, the bound of $\nicefrac{1}{4\sqrt{N}}$, which we prove for composable security, has a tight scaling as implicitly shown in the appendix (cf. Lemma \ref{lemma:secOfProto}). This tight scaling and the tightness of the naive approach (cf. Theorem \ref{theorem:optimalFidAttack} and Lemma \ref{lemma:secOfProto}) for fixed round numbers pose a crucial problem for quantum state verification, especially in the context of composed protocols, even against rather limited attacks, such as i.i.d. attacks. 
        While we restrict the protocol type in the main part of this work to a fixed number of rounds, we prove the same bounds for a probabilistic number of rounds in the appendix (cf. Theorems \ref{theorem:FidGen} and \ref{theorem:composGen}), further closing loopholes to circumvent this no-go result. Importantly, this protocol type was not affected by previous works, and intuitive approaches to break verification are only available for a few combinations of distributions for the number of rounds and output round. 

        Our proofs furthermore shed light on an interesting trade-off in the attack between the acceptance probability and the deviation from the target state. While the intuitive attack discussed in Section \ref{sec:ng-non-iid} yields a constant fidelity of $0$ accepted with a probability $\nicefrac{1}{N+1}$ for a uniform choice of the output round, the acceptance probability in the i.i.d. attacks converges to a constant and the trace distance decreases or fidelity increases in the number of rounds. For example, we find for composable security that the acceptance probability for a protocol that projects onto a pure target state is lower bounded by $\nicefrac{1}{2}$ and the trace distance between the sent state and the target state is $\nicefrac{1}{2\sqrt{N}}$. Hence, even with $10^4$ rounds, the probability of accepting a state that is (at least) $\nicefrac{1}{200}$ apart from the target state is at least $\nicefrac{1}{2}$.
        
        Nevertheless, we note that quantum state verification is not a lost cause; although the cut-and-choose approach cannot yield desirable security using composition theorems or in a stand-alone fashion, one might prove security of a composition with cut-and-choose quantum state verification in a non-modular fashion. Further, other verification mechanisms, such as error detection, might allow for better security but fall out of the scope of this work, which was about the direct cut-and-choose approach for verification. Finally, further research is needed to evaluate the potential of quantum state verification in general, and we need to find out where techniques similar to ours provide further limitations and where positive results with negligible distinguishing advantage can be found. 

 \section{Acknowledgements}
 F. W., Z.C. and A. P. acknowledge support from the DFG via the Emmy Noether grant No. 41829458 and the Hector Fellow Academy. This work was funded by the European Union's Horizon Europe research and innovation programme under grant agreement No. 101102140 -- QIA Phase 1. This project was financially supported by BERLIN QUANTUM, an initiative endowed by the Innovation Promotion Fund of the city of Berlin. M. K. is supported by EPSRC grant EP/V040944/1 Resources in Computation.
\bibliography{Bibliography}
\newpage
\appendix
\section{A more general protocol type}
While we already provide our no-go result for a rather general type of protocols, one might want to go a few steps further in generalization. The most important aspect that is missing in the main matter is a probabilistic round number: one might choose not to fix the round number in advance but during the protocol. While we do not model the communication between the source and the clients in detail, we introduce the following parameters:
\begin{itemize}
    \item $\Omega:\N \rightarrow [0,1]$ is the probability distribution that governs the number of verification rounds. The total number of rounds is one plus the number of verification rounds. 
    \item $(\omega_n:\{0,...,n\}\rightarrow[0,1])_{n\in \N_+}$ is a sequence of probability distributions for the output round.
    \item $\mu_{n,i}(0)\in Pos(\X^{\otimes n})$ is the measurement operator associated with acceptance in the protocol when the protocol uses $n$ rounds and outputs in round $1\leq i\leq n$.
\end{itemize}
We now derive the bound for this protocol for stand-alone security.
\begin{theorem}\label{theorem:FidGen}
  Let $\pi = (\pi_C,\pi_S)$ be a protocol as described above. If $\pi$ is $\varepsilon_H$-correct and $\varepsilon_D$ secure according definition \ref{def:StAl_sec}, it holds
        \begin{align*}
            \varepsilon_H+\varepsilon_D\geq \frac{1}{7N},
        \end{align*}
        where $N$ is the expected number of verification rounds.
    \end{theorem}
    \proof
    The proof follows the same ideas as the proof of Theorem \ref{theorem:FidNGsimple} but uses Jensen's inequality to discard the probabilistic round number. We first set 
    \begin{align*}
        p^H_{n,i} &\coloneqq \braket{\mu_{n,i}(0)}{\phi^{\otimes n}}\\
        p^D_{n,i} &\coloneqq \braket{\mu_{n,i}(0)}{\psi^{\otimes n}}.
    \end{align*}
    where $\psi$ is the state the dishonest source sends for each round.
    Now, we can define 
    \begin{align*}
        p^H_A\coloneqq\sum_{n=0}^{\infty}\Omega(n)\sum_{i=0}^n\omega_n(i) p^H_{n,i}\\
        p^D_A\coloneqq\sum_{n=0}^{\infty}\Omega(n)\sum_{i=0}^n\omega_n(i) p^D_{n,i}
    \end{align*}
    and find 
    \begin{align*}
        \varepsilon_H&\geq 1-p^H_A\\
        \varepsilon_D&\geq p^D_A(1-F(\phi,\psi))
    \end{align*}
    following the same manipulations as in the proof of Theorem \ref{theorem:FidNGsimple} with an i.i.d. attack using $\psi$.\\
    Hence, we find for the sum of $\varepsilon_H$ and $\varepsilon_D$
    \begin{align*}
        \varepsilon_H+\varepsilon_D&\geq 1-p^H_A+p^D_A(1-F(\phi,\psi)) \geq (1-F(\phi,\psi))\left(1-\left|p^H_A-p^D_A\right|\right).
    \end{align*}
    Now the differences between the protocol types come into play, as $|p^H_A-p^D_A|$ is the expected distinguishing advantage with a probabilistic number of states, i.e., using \eqref{eq:Multi_copy_dist}:
    \begin{align*}
        |p^H_A-p^D_A| &= \sum_{n=0}^{\infty}\Omega(n)\sum_{i=0}^n\omega_n(i) |p^H_{n,i}-p^D_{n,i}| \leq \sum_{n=0}^{\infty}\Omega(n)\frac{1}{2}\left\|\phi^{\otimes {n}}-\psi^{\otimes {n}}\right\|\leq \sum_{n=0}^{\infty}\Omega(n)\sqrt{1-F(\phi,\psi)^n}
    \end{align*}
    As $f(n) = \sqrt{1-a^n}$ is concave for $0\leq a\leq 1$ (cf. Lemma \ref{Lemma:Concave}) we can use Jensen's inequality and find
    \begin{align*}
        |p^H_A-p^D_A| \leq \sqrt{1-F(\phi,\psi)^N},
    \end{align*}
    where $N$ is the expected number of verifications.
    Combined with the above, this means
    \begin{align*}
        \varepsilon_H+\varepsilon_D\geq (1-F(\phi,\psi))\left(1-\sqrt{1-F(\phi,\psi)^N}\right).
    \end{align*}
    Now we set $\tau \coloneqq (1-F(\phi,\psi))$, i.e. $F(\phi,\psi) = (1-\tau)$:
    \begin{align*}
        \varepsilon_H+\varepsilon_D\geq \tau\left(1-\sqrt{1-(1-\tau)^{N}}
        \right).
    \end{align*}
    We choose $\tau = \nicefrac{\alpha}{N}$ and get
    \begin{align*}
        \varepsilon_H+\varepsilon_D\geq \frac{\alpha}{N}\left(1-\sqrt{\alpha}\right),
    \end{align*}
    which is optimal for $\alpha=\nicefrac{4}{9}$, as was shown in the proof of Theorem \ref{theorem:FidNGsimple}, which gives
    \begin{align*}
        \varepsilon_H+\varepsilon_D\geq \frac{4}{27N} \geq \frac{1}{7N}
    \end{align*}\qed
    
We also find the same result as for the simpler protocol type with regard to composable security.
\begin{theorem}\label{theorem:composGen}
    Let $\pi = (\pi_C,\pi_S)$ be a protocol as above. If 
    \begin{align*}
        \frac{1}{2}\left\|\rho_H-(\phi\oplus 0)\right\|_1\leq \varepsilon_H
    \end{align*}
    and
    \begin{align*}
        \min_{p\in[0,1]}\frac{1}{2}\left\|\rho_D-(p\phi\oplus (1-p)\right\|_1\leq \varepsilon_D.
    \end{align*}
    It holds $\varepsilon_H+\varepsilon_D\geq \frac{\sqrt{\eta_1}}{4\sqrt{N}}$ if $\phi$ is mixed and $\eta_1$ is the largest eigenvalue of $\phi$ and $\varepsilon_H+\varepsilon_D\geq \frac{1}{4\sqrt{N}}$ if $\phi$ is pure.
    \end{theorem}
    \proof As in the previous theorem, we can use the result for a fixed round number (Theorem \ref{theorem:composSimple}) by using Jensen's inequality. Again we denote 
        \begin{align*}
        p^H_{n,i} &\coloneqq \braket{\mu_{n,i}(0)}{\phi^{\otimes n}}\\
        p^D_{n,i} &\coloneqq \braket{\mu_{n,i}(0)}{\psi^{\otimes n}}\\
        p^H_A &\coloneqq \sum_{n=0}^{\infty}\Omega(n)\sum_{i=0}^np^H_A\\
        p^D_A &\coloneqq \sum_{n=0}^{\infty}\Omega(n)\sum_{i=0}^np^H_A.
    \end{align*}
    Following the same line as for the proof of Theorem \ref{theorem:composSimple}, we find
    \begin{align*}
        \varepsilon_H+\varepsilon_D&\geq \frac{p^D_A}{2} \left\|\psi-\phi\right\|_1 + 1 - p^H_A\geq \frac{1}{2}\left\|\psi-\phi\right\|_1\left(1-\left|p^H_A-p^D_A\right|\right).
    \end{align*}
    Again, we fix $\psi$ now. We write $\phi$ in spectral decomposition $\phi=\sum_{i=1}^d\eta_i\ketbra{\phi_i}$ with $\eta_{i+1}\leq \eta_{i}$ for $1\leq i<d$. We fix $\psi = \eta_1 \ketbra{\chi} + \sum_{i=2}^d\eta_i\ketbra{\phi_i}$ for some pure state $\ket{\chi}$. This implies 
    \begin{align}
        \frac{1}{2}\left\|\phi-\psi\right\|_1 = \eta_1\sqrt{1-|\braket{\phi_1}{\chi}|^2}.
    \end{align}
    In the case of $n$ copies, we utilize the result from the proof of Theorem \ref{theorem:composSimple} and find
    \begin{align*}
        \frac{1}{2}\left\|\phi^{\otimes n}-\psi^{\otimes n}\right\|_1 
        &\leq \sqrt{1-|\braket{\phi_1}{\chi}|^{2\eta_1n}}.
    \end{align*}
    This implies that
    \begin{align*}
        \left|p^H_A-p^D_A\right| &= \sum_{n=0}^{\infty}\Omega(n)\sum_{i=0}^n\omega_n(i)|p^H_{n,i}-p^D_{n,i}|\leq \frac{1}{2}\sum_{n=0}^{\infty}\Omega(n)\left\|\phi^{\otimes n}-\psi^{\otimes n}\right\|_1\\&\leq \sum_{n=0}^{\infty}\Omega(n)\sqrt{1-|\braket{\phi_1}{\chi}|^{2\eta_1n}}.
    \end{align*}
    We use concavity and Jensen's inequality and Lemma \ref{Lemma:Concave} again and find
    \begin{align*}
         \left|p^H_A-p^D_A\right| \leq \sqrt{1-|\braket{\phi_1}{\chi}|^{2\eta_1N}},
    \end{align*}
    where $N$ is the expected number of verification rounds.
    So for $\varepsilon_H+\varepsilon_D$ we find
    \begin{align*}
        \varepsilon_H+\varepsilon_D\geq \eta_1\sqrt{1-|\braket{\phi_1}{\chi}|^2}\left(1-\sqrt{1-|\braket{\phi_1}{\chi}|^{2\eta_1N}}\right).
    \end{align*}
    We know from the proof of the bound for the simpler protocol that we can choose $\ket{\chi}$ such that
    \begin{align*}
        \varepsilon_H+\varepsilon_D\geq\frac{\sqrt{\eta_1}}{4\sqrt{N}}.
    \end{align*}
    In particular, for pure states this gives us
    \begin{align*}
        \varepsilon_H+\varepsilon_D\geq\frac{1}{4\sqrt{N}}.
    \end{align*}\qed

\section{Tightness of scaling}
In order to investigate how tight the lower bounds that we provide are, we analyze the correctness and security of a protocol of the type shown in Protocol \ref{proto:simple}, in which the target state is pure, every round has the same probability to be the output round, and the measurement is always a projection onto $N$ copies of the target state. We find that this choice of parameters yields a protocol which is perfectly correct in both, stand-alone and composable security, $\nicefrac{1}{N+1}$-secure in stand-alone and $\nicefrac{2}{\sqrt{N+1}}$-secure in composable security, respectively.

These results are in line with the i.i.d. bound of $\nicefrac{1}{7N}$ given for stand-alone security in Theorem \ref{theorem:FidNGsimple},  the i.i.d. bound of $\nicefrac{1}{4\sqrt{N}}$ given for composable security in Theorem \ref{theorem:composSimple} and the bound for stand-alone security of $\nicefrac{1}{N+1}$ given by the combination of Lemma \ref{lemma:equal}, Lemma \ref{lemma:genlowbound} and Theorem \ref{theorem:optimalFidAttack} in Section \ref{sec:ng-non-iid}. The last of these three bounds was shown to be optimal in the class of separable state attacks. Indeed, the following lemma proves that it is optimal, since it is tight, in the class of all attacks for the protocol considered in this section.
For the two i.i.d. bounds we can use $\nicefrac{1}{N+1}\leq\nicefrac{1}{N}$ and $\nicefrac{2}{\sqrt{N+1}}\leq\nicefrac{2}{\sqrt{N}}$, which implies that the protocol is also $\nicefrac{1}{N}$-secure in stand-alone and $\nicefrac{2}{\sqrt{N}}$-secure in composable security. Hence, the lower bounds for $\varepsilon_H+\varepsilon_D$ with the i.i.d. restriction for the attacker we presented before exhibit the same scaling in $N$ as the upper bounds for the same quantities for this specific protocol without restrictions on the attacks. Therefore, the following lemma implies that the scaling of the i.i.d. bounds is tight and no attack can violate security for all choices of parameters in Protocol \ref{proto:simple} with a better scaling in $N$. Nevertheless, there might be tighter general bounds with the same scaling or bounds with better scaling for specific choices of parameters in Protocol \ref{proto:simple}.

\begin{lemma}\label{lemma:secOfProto}
    Let $\pi=(\pi_C,\pi_S)$ be a protocol as described in Protocol \ref{proto:simple} where the parameters are chosen so that 
        \begin{itemize}
            \item $\phi = \ketbra{\phi_0}\in D(\X)$ is the pure target state,
            \item for all $i\in\{1,..., N+1\}$ it holds $\omega(i) = \nicefrac{1}{N+1}$, i.e., every round has the same probability to be the output round,
            \item and for all $i\in\{1,...,N+1\}$ the clients use $\mu_i(0) = \ketbra{\phi^{\otimes N}_0}$.
        \end{itemize}
    This implies 
    \begin{enumerate}
        \item $\pi$ is $0$-correct and $\nicefrac{1}{N+1}$-secure with respect to Definition \ref{def:StAl_sec}.
        \item $\pi$ is $0$-composable correct with respect to $\sharp_S(\S_{\phi}^{QSV})$ and $\nicefrac{2}{\sqrt{N+1}}$-composable secure with respect to $\S_{\phi}^{QSV}$.
    \end{enumerate}
\end{lemma}
\proof While correctness follows immediately, we use a symmetry argument for security since the clients choose the output round randomly and bound a sum of diagonal elements by the trace. Finally, we find the composable security from a known result proven in \cite{colisson2024graph}.

The correctness in the first statement follows from $\braket{\ketbra{\phi_0}^{\otimes N}} = |\braket{\phi_0}|^N = 1$.
Hence, the protocol is perfectly correct with respect to Definition \ref{def:StAl_sec}, since the probability to reject the behavior of an honest source is $0$..\\
If the source is dishonest, we assume that it sends a state $\psi\in D(\X^{\otimes N+1})$, which might be entangled across the rounds, consider the completion of $\ket{\phi_0}$ to an orthonormal basis $\{\ket{\phi_i}\}_{i=0}^{\rm{dim}(\X)-1}$ of $\X$ and define $A_{\ell},\ B_{\ell}\in\mathcal{U}(\X^{\otimes N+1})$ for $\ell\in\{1,...,N+1\}$ as 
\begin{align*}
    \ell\in\{1,...,N\}:\ A_{\ell}&\coloneqq \sum_{i,j=0}^{\rm{dim}(\X)-1} \left(\mathbbm{1}_{\X}^{\otimes \ell-1}\otimes \ket{\phi_i} \otimes\ket{\phi_j} \otimes  \mathbbm{1}_{\X}^{\otimes N-\ell}\right)\left(\mathbbm{1}_{\X}^{\otimes \ell-1}\otimes \bra{\phi_j} \otimes\bra{\phi_i} \otimes  \mathbbm{1}_{\X}^{\otimes N-\ell}\right),\\
    A_{N+1}&\coloneqq \mathbbm{1}_{\X}^{\otimes N+1}\\
    \ell\in\{1,...,N+1\}:\ B_{\ell}&\coloneqq A_{N}A_{N-1}...A_{\ell+1}A_{\ell},
\end{align*}
hence, $A_{\ell}$ is the unitary that exchanges the $\ell$-th and the $(\ell+1)$-th register, and $B_{\ell}$ pushes the $\ell$-th register to the last position. Equipped with these definitions, we find that if the source is dishonest, the average state is given by
\begin{align*}
    \rho_D =& \left(\bra{\phi_0}^{\otimes N}\otimes \mathbbm{1}_{\X}\right) \left(\frac{1}{N+1}\sum_{\ell=1}^{N+1}B_{\ell}\psi B_{\ell}\D\right)\left(\ket{\phi_0}^{\otimes N}\otimes \mathbbm{1}_{\X}\right)\ \oplus \\&
    \left(1-\Tr(\left(\bra{\phi_0}^{\otimes N}\otimes \mathbbm{1}_{\X}\right) \left(\frac{1}{N+1}\sum_{\ell=1}^{N+1}B_{\ell}\psi B_{\ell}\D\right)\left(\ket{\phi_0}^{\otimes N}\otimes \mathbbm{1}_{\X}\right))\right),
\end{align*}
where we again used $\braket{\ketbra{\phi_0}^{\otimes N}}{\chi} = \bra{\phi_0}^{\otimes N}\chi\ket{\phi_0}^{\otimes N}$.
Using the behavior of the fidelity under direct sums and the optimal ideal acceptance probability $p$ we find
\begin{align*}
    \max_{p\in[0,1]}F(\rho_D,p\phi\oplus (1-p)) &= F\left(\phi,\left(\bra{\phi_0}^{\otimes N}\otimes \mathbbm{1}_{\X}\right) \left(\frac{1}{N+1}\sum_{\ell=1}^{N+1}B_{\ell}\psi B_{\ell}\D\right)\left(\ket{\phi_0}^{\otimes N}\otimes \mathbbm{1}_{\X}\right)\right) + \\
    &1-\Tr(\left(\bra{\phi_0}^{\otimes N}\otimes \mathbbm{1}_{\X}\right) \left(\frac{1}{N+1}\sum_{\ell=1}^{N+1}B_{\ell}\psi B_{\ell}\D\right)\left(\ket{\phi_0}^{\otimes N}\otimes \mathbbm{1}_{\X}\right)).
\end{align*}
Next, we use that $F(\rho,\ketbra{\gamma}) =\bra{\gamma}\rho\ket{\gamma}$, express the trace with the basis $\{\ket{\phi_0}\}_{i=0}^{\rm{dim}(\X)-1}$ and apply $B_{\ell}$ on the vectors instead of the operator and find
\begin{align*}
    &\max_{p\in[0,1]}F(\rho_D,p\phi\oplus (1-p)) = \bra{\phi_0}^{\otimes N+1} \left(\frac{1}{N+1}\sum_{\ell=1}^{N+1}B_{\ell}\psi B_{\ell}\D\right)\ket{\phi_0}^{\otimes N+1}  \\
    +&1-\sum_{i=0}^{\rm{dim}(\X)-1}\left(\bra{\phi_0}^{\otimes N}\otimes \bra{\phi_i}\right) \left(\frac{1}{N+1}\sum_{\ell=1}^{N+1}B_{\ell}\psi B_{\ell}\D\right)\left(\ket{\phi_0}^{\otimes N}\otimes \ket{\phi_i}\right)\\
    =& 1-\sum_{i=1}^{\rm{dim}(\X)-1}\left(\bra{\phi_0}^{\otimes N}\otimes \bra{\phi_i}\right) \left(\frac{1}{N+1}\sum_{\ell=1}^{N+1}B_{\ell}\psi B_{\ell}\D\right)\left(\ket{\phi_0}^{\otimes N}\otimes \ket{\phi_i}\right)\\
    =& 1-\frac{1}{N+1}\sum_{i=1}^{\rm{dim}(\X)-1}\sum_{\ell=1}^{N+1} \left(\bra{\phi_0}^{\otimes \ell-1}\otimes \bra{\phi_i}\otimes  \bra{\phi_0}^{\otimes N-\ell+1}\right)\psi  \left(\ket{\phi_0}^{\otimes \ell-1}\otimes \ket{\phi_i}\otimes  \ket{\phi_0}^{\otimes N-\ell+1}\right).
\end{align*}
To prove the first claim of the lemma, it is now sufficient to recognize that the sum in the above equation adds up some of the diagonal elements of $\psi$. However, since $\psi\in D(\X^{N+1})$, this sum cannot exceed the trace of $\psi$, which is $1$, i.e.
\begin{align*}
    \max_{p\in[0,1]}&F(\rho_D,p\phi\oplus (1-p)) \geq 1-\frac{1}{N+1}.
\end{align*}
Composable correctness follows immediately from the impossibility of rejecting the behavior of the source if it is honest. For the proof of composable security, we refer to \cite{colisson2024graph}, which states (with our definition of fidelity) that 
\begin{align*}
    \max_{p\in[0,1]}&F(\rho_D,p\phi\oplus (1-p)) \geq \sqrt{1-\kappa},
\end{align*}
implies $2\sqrt{2\kappa-\kappa^2}$- composable security. Note that if all clients are honest, the necessity that $\phi$ is a graph state does not apply, and the ideal resources in \cite{colisson2024graph} are identical to $\S_{\phi}^{QSV}$. Since it holds
\begin{align*}
    2\sqrt{2\kappa-\kappa^2} = 2\sqrt{1-(1-\kappa)^2},
\end{align*}
and we just proved $\max_{p\in[0,1]}F(\rho_D,p\phi\oplus (1-p)) \geq 1-\frac{1}{N+1}$ we can set $(1-\kappa)^2 = 1-\frac{1}{N+1}$, which proves $\nicefrac{2}{\sqrt{N+1}}$ composable security.\qed

\end{document}